# High Performance Broadband Photodetection Based on Graphene – MoS$_{2x}$Se$_{2(1-x)}$ Alloy Engineered Phototransistors

Shubhrasish Mukherjee[1], Didhiti Bhattacharya[1], Samit Kumar Ray[*1,2] and Atindra Nath Pal[*1]

*[1] S. N. Bose National Center for Basic Science, Sector III, Block JD, Salt Lake, Kolkata – 700106*

*[2] Indian Institute of Technology Kharagpur, 721302, West Bengal, India*

Email: physkr@phy.iitkgp.ac.in, atin@bose.res.in

## Abstract

The concept of alloy engineering has emerged as a viable technique towards tuning the bandgap as well as engineering the defect levels in two-dimensional transition metal dichalcognides (TMDC). Possibility to synthesize these ultrathin TMDC materials through chemical route has opened realistic possibilities to fabricate hybrid multi-functional devices. By synthesizing nanosheets with different composites of MoS$_{2x}$Se$_{2(1-x)}$ (x = 0 to 1) using simple chemical methods, we systematically investigate the photo response properties of three terminal hybrid devices by decorating large area graphene with these nanosheets (x = 0, 0.5, 1) in 2D-2D configurations. Among them, graphene-MoSSe hybrid phototransistor exhibits superior optoelectronic properties than its binary counterparts. The device exhibits extremely high photoresponsivity (>$10^4$ A/W), low noise equivalent power (~$10^{-14}$ W/Hz$^{0.5}$), higher specific detectivity (~ $10^{11}$ Jones) in the wide UV-NIR (365-810 nm) range with excellent gate tunability. The broadband light absorption of MoSSe, ultrafast charge transport in graphene, along with controllable defect engineering in MoSSe makes this device extremely attractive. Our work demonstrates the large area scalability with wafer-scale production of MoS$_{2x}$Se$_{2(1-x)}$ alloys, having important implication towards facile and scalable fabrication of high-performance optoelectronic devices and providing important insights into the fundamental interactions between van-der-Waals materials.

## Introduction

Broadband photodetectors are of particular importance due to their applicability in the large dynamic range of the electromagnetic spectrum, improving versatility in wide application fields, such as flexible image sensing, surveillance, communication and health monitoring[1,2]. Two dimensional (2D) layered semiconductors, in particular, transition metal dichalcognides (TMDC)[3] have emerged as interesting materials because of their controllable band gap (1-2 eV) and strong light-matter interactions[4]. Various strategies like strain engineering[5], varying the layer numbers in the pristine materials are employed to widen the effective wavelength range. However, most of the processes are either difficult to reproduce or may severely affect the performance of the phototransistors like responsivity, response time etc[6]. Two dimensional heterostructures[7] have emerged as a powerful route towards creating high responsivity phototransistor by incorporating two different 2D layered materials in close proximity through van der Waals interaction[8]. Additionally, the possibility of creating their nanostructures using rather simple chemical routes has shown to widen the absorption band width and hence, can be useful for broadband photodetection[9]. Among various possibilities, graphene based heterostructure has shown immense prospect. In spite of having superior electronic properties[10,11] like high charge carrier mobilities, broad spectral bandwidth, ultra large specific surface area etc., it has limitations in the optoelectronic applications due to its low light absorption and gapless nature[12]. One simple strategy to overcome this shortcoming is to create a new device structure by incorporating some light absorbing (Si quantum dots[13], PbS[14], ZnO[15,16] etc) materials into graphene. In this typical hybrid structure graphene is used for carrier transport channel and interaction between the photosensitive material and graphene is the prime factor for the ultrasensitive photodetection. Being layered semiconductors, some members of the transition metal dichalcogenide[3] (TMDC) family ($MX_2$ (M = Mo, W; X = S, Se) are natural partners of graphene for optically active heterostructures[8]. However, these binary TMDCs suffer from intrinsic defects like chalcogen vacancies, which strongly affect their intrinsic electronic and optical properties leading to localized deep-level defect states (DLDSs)[17]. Alloying is considered to be the potential solution to overcome such difficulties as it offers lower deep level densities[18], conversion of deep to shallow levels of defects and also better thermal stability[19]. Hence, for the broad applications in integrated devices, growth of these bandgap engineered 2D ternary alloys may provide realistic solution to improve the device performance. It is already reported that $MoS_{2x}Se_{2(1-x)}$ ternary alloys can successfully be prepared[20], but the requirement of high temperature, specific atmospheric conditions etc. to grow these materials hinder their application related performances[21].

This paper presents a systematic study of graphene-$MoS_{2x}Se_{2(1-x)}$ hybrid phototransistors with x varying from 0 to 1. Using a simple hydrothermal synthesis followed by ultrasonication, composition-tunable ternary $MoS_{2x}Se_{2(1-x)}$ alloy nanosheets are synthesized and their structural and optical investigations demonstrate that these 2D ternary alloys are of highly crystalline and exhibit composition-dependent band edge emission. Comparing to their binary counterparts, the higher photoresponsivity and faster response time in graphene-MoSSe phototransistor imply the reduced density of deep level defects. This fabricated hybrid phototransistor using the optimized alloy exhibits a gate tunable photoresponsivity (R) higher than $10^4$ A/W and a noise equivalent power as low as ~$10^{-14}$ W/$Hz^{0.5}$ in the broad spectral region 365-810 nm. Also considering the low frequency 1/f noise measurement, the specific detectivity ($D^*$) of the device reaches ~$10^{11}$ Jones, which is higher or comparable to the previously reported results[22,23]. Such a stable, highly sensitive phototransistor demonstrated here has the potential to be used in next generation multifunctional devices.

**Results and discussions**

$MoS_{2x}Se_{2(1-x)}$ (x = 0 to 1) nanosheets are synthesized by using sono-chemical assisted hydrothermal technique[24]. In comparison to other methods, this hydrothermal technique is an easy, cost-effective and eco-friendly technique to synthesize 2D materials and their alloys with excellent composition tunability[25]. All the compositions of $MoS_{2x}Se_{2(1-x)}$ alloy nanosheets are investigated by using scanning electron microscopy (SEM) equipped with energy-dispersive X-ray (EDAX) spectroscopy (**Figure S1**). The signature of Mo, S, Se can be observed in EDAX spectrum of the ternary alloys and their homogeneous distribution in the typical alloy nanosheets (for x= 0, 0.5, 1) is clearly evidenced by the elemental mapping images (**Figure S2**). Transmission electron microscopy (TEM) and atomic force microscopy (AFM) images (**Figure S3a and S3b,** respectively) confirm the formation of few layered nanosheets of $MoS_{2x}Se_{2(1-x)}$ alloys with varying composition. These 2D nanosheets are well dispersed in IPA and form a stable colloidal dispersion (Inset, **Figure S3a**).

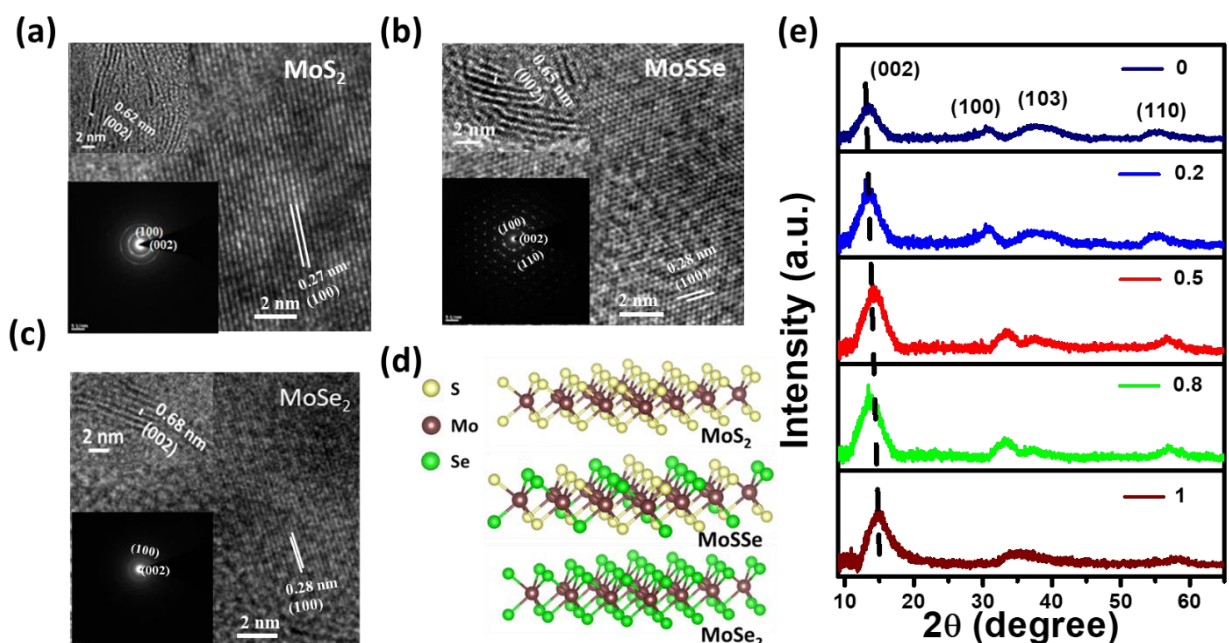

**Figure 1.** Microstructural characteristics of $MoS_{2x}Se_{2(1-x)}$ alloys. High-resolution TEM images of a) $MoS_2$, b) MoSSe and c) $MoSe_2$ nanosheets with corresponding interlayer spacing of (002) plane (top Inset) and SAED patterns (bottom Inset). d) Schematic structures of monolayer $MoS_2$, MoSSe and $MoSe_2$ nanosheets. e) XRD patterns of different composites of $MoS_{2x}Se_{2(1-x)}$ alloys with varying x.

To understand the crystalline character, high-resolution transmission electron microscopy (HRTEM) images and the selected area electron diffraction (SAED) patterns (bottom Inset) of the typical MoSSe (x = 0.5) ternary alloy along with the binary counterparts $MoS_2$ (x = 1) and $MoSe_2$ (x = 0) are shown in **Figure 1a, b, c,** respectively. Based on the HRTEM images, the lattice spacing of MoSSe alloy is found to be 0.28 nm, which is in good agreement with the formation of (100) plane[26] (**Figure 1b**). The SAED pattern of a typical MoSSe nanosheet shows bright diffraction spots with high quality hexagonal symmetry structure, with the outer and inner spots indexed as the (110) and (100) planes of 2H-MoSSe, respectively[27]. With increasing Se content, the interlayer spacing for (002) plane (top Inset) is found to be increased (for $MoS_2$~0.62 nm, MoSSe~0.65 nm, $MoSe_2$~0.68 nm) due to the larger atomic radius of Se than S[28]. Pristine monolayer $MoS_2$ ($MoSe_2$) consists of three sublayers, one layer of Mo atom sandwiched by two planes of S (Se) atoms. In the hydrothermal synthesis, the sulpho-selenide compositions S and Se's are distributed randomly, keeping the Mo atomic sites at the centre of the tetragon and maintain the hexagonal crystal symmetry, as depicted schematically in **Figure 1d**. The XRD patterns of all these ternary alloys (**Figure 1e**) along with the two binary counterparts exhibit diffraction peaks which are well consistent with the previous reports[29,30]. Decreasing the S content leads to the continuous downshift of (002) diffraction peak in $MoS_{2x}Se_{2(1-x)}$ ternary alloys. This is in accordance with the gradual expansion of lattice structure due to substitution of S atoms by larger radii Se atoms, which implies that the mixing occurs in atomic level and the obtained results from the XRD analysis agree well with the HRTEM observations.

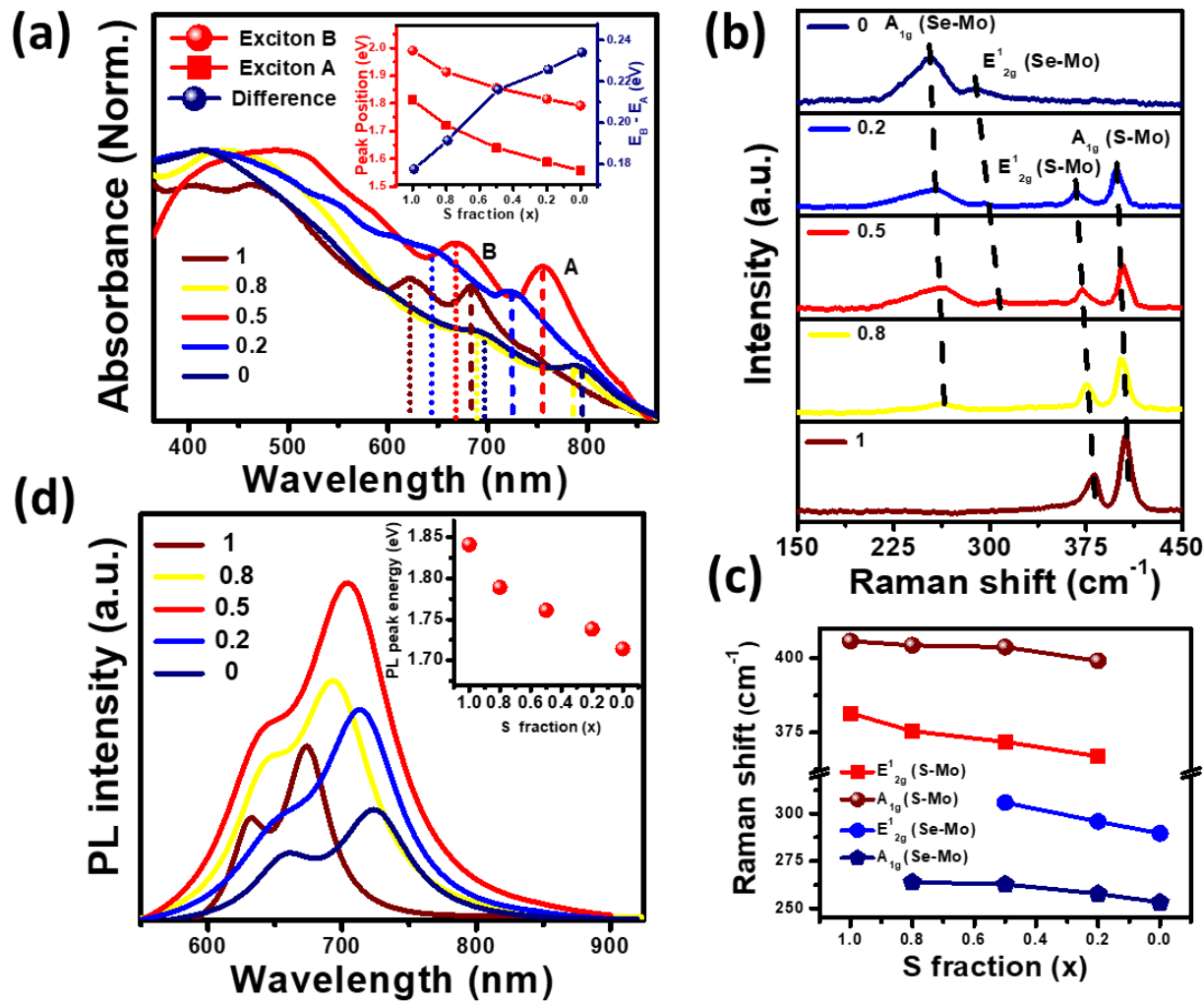

**Figure 2.** Spectroscopic characterizations of $MoS_{2x}Se_{2(1-x)}$ alloys a) Normalized absorbance spectra with different x values, energy of excitonic A and B peaks (represented with dashed lines for exciton 'A' and dotted lines as exciton 'B') and their separation as a function S fraction (Inset). b) Raman spectra of the alloys excited with 532 nm laser. c) Variation of S-Mo and Se-Mo related Raman mode shift with S fraction. d) Composition dependent PL spectra of the alloys and the emission peak position as a function of x (Inset).

The spectroscopic properties of the few layered as-synthesized ternary $MoS_{2x}Se_{2(1-x)}$ alloys are investigated by using UV-Vis absorption spectroscopy (**Figure 2a**). All the alloy composites (x = 0 to 1) show two prominent peaks, designated as exciton 'A' and 'B' peaks. These peaks originate from the direct excitonic transitions at the K and K´ points of the first Brillouin zone and the spin-orbit coupled valance bands[4,31]. It is observed that, with decreasing S (increasing Se) content both the excitonic peaks (exciton 'A' and 'B') shift to the lower energies (**Inset Figure 2a**). Also, the energy difference between these two peaks ($E_B-E_A$) increases monotonically (from 0.17 eV to 0.23 eV) with decreasing S content (from 1 to 0), which can be explained by the enhanced spin-orbit coupling strength with increasing Se content[32].

Composition tunable vibrational modes can be studied from the Raman spectra of $MoS_{2x}Se_{2(1-x)}$ alloys (**Figure 2b**). All these ternary alloys exhibit 4 different vibrational modes which can be assigned as $A_{1g}$ (Se-Mo), $E^1_{2g}$ (Se-Mo), $A_{1g}$ (S-Mo) & $E^1_{2g}$ (S-Mo). All of these molecular vibrational modes are in good agreement with the gradual tuning of the S or Se component in $MoS_{2x}Se_{2(1-x)}$ alloys. The modes related to the Se-Mo vibration i.e., $A_{1g}$ (Se-Mo) and $E^1_{2g}$ (Se-Mo), remain absent in $MoS_2$ (x = 1) and start to emerge with introducing Se content. Similarly, the intensity of the S−Mo related modes ($E^1_{2g}$(S−Mo), $A_{1g}$(S−Mo)) gradually decreases and disappear entirely in the binary $MoSe_2$ compound. All the vibrational modes shift to a lower frequency with decrease of S content (increase Se content), as shown in **Figure 2c**. With the decrease of S (increase of Se) fraction, the interaction of Se and S atoms would soften the S-Mo related modes and therefore the modes related to Se-Mo shift to lower frequencies due to the decrease in S content[33].

The photoluminescence (PL) spectra, recorded by an excitation with 532 nm laser of $MoS_{2x}Se_{2(1-x)}$ alloys reveal (**Figure 2d**) that the emission spectrum can be modulated by varying the S and Se content. It is observed that the PL intensity of MoSSe alloy is greatly enhanced compared to the others. With the introduction of Se, the PL emission peak shows a red shift starting from 673 nm for pure $MoS_2$ to 725 nm for pure $MoSe_2$. These band edge excitonic transitions are attributed to the valance band splitting which is affected by the enhanced spin-orbit coupling due to decreasing S (increasing Se content)[34]. So, by tuning the Se and S component in the ternary alloys, the optical band gap can be modulated as it is observed from the emission peak shift in PL measurements (**Inset, Figure 2d**).

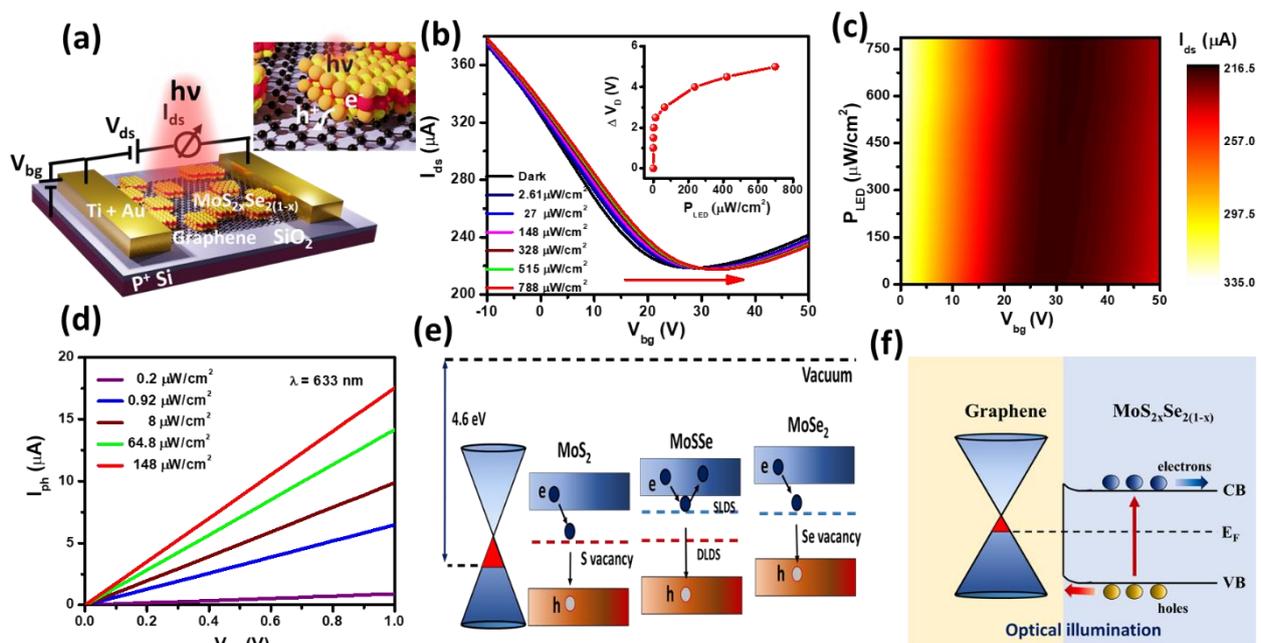

**Figure 3.** Characteristics of graphene-MoSSe heterostructure device a) Schematic of the hybrid device where CVD grown graphene on $Si/SiO_2$ is coated with $MoS_{2x}Se_{2(1-x)}$ alloy nanosheets. b) Drain current ($I_{ds}$) as a function of back gate voltage ($V_{bg}$) for graphene-MoSSe alloy hybrid phototransistor under different illumination powers at $V_{ds}$ = 0.5 V, λ = 633 nm. Shift of Dirac point (Δ $V_D$) as a function of illumination power ($P_{LED}$) (Inset). c) Two dimensional (2D) colour plot of drain current ($I_{ds}$) as a function of optical power and gate voltage. d) Photocurrent ($I_{ph}$) as a function of source-drain bias voltage ($V_{ds}$) for different optical powers at $V_{bg}$ = 0 V, λ = 633 nm. e) Comparative energy band diagram for graphene, $MoS_2$, MoSSe and $MoSe_2$ along with the schematic localized defect states. f) Suggested energy level diagram of the heterostructure illustrating the photogenerated electron- hole separation at the interface.

Based on the optical emission and absorption characteristics of the ternary alloys, we have focused on their optoelectronic behavior using graphene-based hybrid back-gated phototransistor. **Figure 3a** shows the schematic of the device based on monolayer graphene and 2D TMDC alloys. In this typical device structure, large area chemical vapor deposited (CVD) graphene acts as the channel material between two Ti/Au electrodes on top of a Si/SiO$_2$ substrate (300 nm, p$^+$ doped). Synthesized TMDC ternary alloys (MoS$_{2x}$Se$_{2(1-x)}$) are drop-casted on top of the graphene to form the hybrid devices (see **Figure S4a**, **b** and **c** for optical images of graphene-MoS$_2$, graphene-MoSSe and graphene-MoSe$_2$ devices respectively). The scanning electron micrograph of the graphene-MoSSe (x = 0.5) heterostructure and the corresponding EDAX spectra are shown in **Figure S4d** and **e**, respectively. Both the Raman (**Figure S4f**) and EDAX spectra (**Figure S4e**) are used as a fingerprint of the presence of graphene and the MoSSe alloy in our measured device, while the steady state PL spectra (**Figure S4g**) is recorded to understand the charge transfer between them. The transfer characteristics (**Figure S5a**) of bare graphene and the heterostructure device (graphene-MoSSe) reveal that the bare CVD grown graphene is a p-doped (Dirac point ($V_D$) > 80 V) and the Dirac point shifts to ~ 28 V due to the presence of MoSSe alloy. This means the p-type doping effect in graphene is weakened due to the effective electron transfer from MoSSe to graphene, which results from the initial work function mismatch between these two materials[35]. The average hole mobility of the hybrid device is found to be ~ 216 cm$^2$/VS (**Figure S5b**). The electron transfer from MoSSe to graphene forms a built-in electric field at the interface which leads to band bending to equilibrate the Fermi level[36] (Schematics in **Figure S5c** and **d**). **Figure 3b** shows the drain-source current ($I_{ds}$) as a function of back gate voltage ($V_{bg}$) under different optical excitation powers at a fixed drain source voltage ($V_{ds}$ = 0.5 V) with $\lambda$ = 633 nm. As the optical intensity increases, the Dirac point ($V_D$) (where the drain-source current ($I_{ds}$) becomes minimum) shifts toward a higher $V_{bg}$, implying the light induced hole doping in the graphene channel. Under illumination, electron-hole pairs are generated at MoSSe alloy and are dissociated at the graphene-MoSSe interface due to the built-in electric field. The photogenerated holes are then transferred to the graphene channel and the electrons remain trapped in MoSSe, inducing the photogating effect through capacitive coupling[14]. The photoinduced shift of $V_D$ with illumination intensity for $\lambda$ = 633 nm and $V_{ds}$ = 0.5 V is shown in **Figure 3b (inset)**. The shift is more evident in **Figure 3c** which represents the 2D colour plot of $I_{ds}$ as a function of $V_{bg}$ and the illumination intensity ($P_{LED}$) of $\lambda$ = 633 nm. **Figure 3d** shows the variation of photocurrent ($I_{ph}$) with $V_{ds}$ at zero gate bias for $\lambda$ = 633 nm. We see that $I_{ph}$ grows linearly with $V_{ds}$ and also, increases monotonically with the increase in optical intensity. With increasing optical intensity, more numbers of charge carriers are generated in the 2D TMDC alloy inducing a stronger photogating effect and higher photocurrents[37,38]. **Figure 3e** illustrates the comparative energy band diagram of graphene and MoS$_{2x}$Se$_{2(1-x)}$ alloys with different compositions (x = 0, 0.5, 1) along with their deep level defects (DLDS) and shallow level defects (SLDS). In the binary layered TMDC materials (like MoS$_2$, MoSe$_2$ etc.), the unintended presence of chalcogen vacancies are available in plenty leading to the formation of localized DLDSs within the bandgap, which act as recombination trap centres for the photoexcited charge carriers and greatly affect the device performance[20]. However, the formation of ternary MoSSe alloy improves thermodynamic stability due to its low formation energy[19]. Also, the energy levels are re-localized and the intrinsic defect states are superimposed because of the coexistence of S and Se vacancies[39]. So, the harmful deep level defects in MoS$_2$ and MoSe$_2$ become shallow level defects in MoSSe ternary alloy[40]. The model that illustrates the physical mechanism behind the photocurrent generation in graphene-MoSSe heterostructure device is represented in **Figure 3f**. Here, the illuminated light is absorbed by the TMDC alloy and subsequently the electron-hole pairs are generated, as shown in the **Figure 3f**. The photogenerated holes are then transferred to the graphene channel due to the upward band bending producing a strong photoresponse as discussed earlier[41,42].

The photoresponsivity (R)[43], $R = \frac{I_{ph}}{P_{LED}}$, of the hybrid phototransistor as a function of optical intensities ($P_{LED}$) are shown in **Figure 4a**, where all the responsivity measurements are performed under identical experimental conditions at $V_{ds} = 0.5$ V, $V_D - V_{bg} = 15$ V with the varying wavelength of 365-810 nm. The expected inverse relationship between R and $P_{LED}$ is also observed here and can be explained by the saturation of the photocurrent due to the lowering of the built-in field at the interface with an increase in the number of charge carriers[44,45].

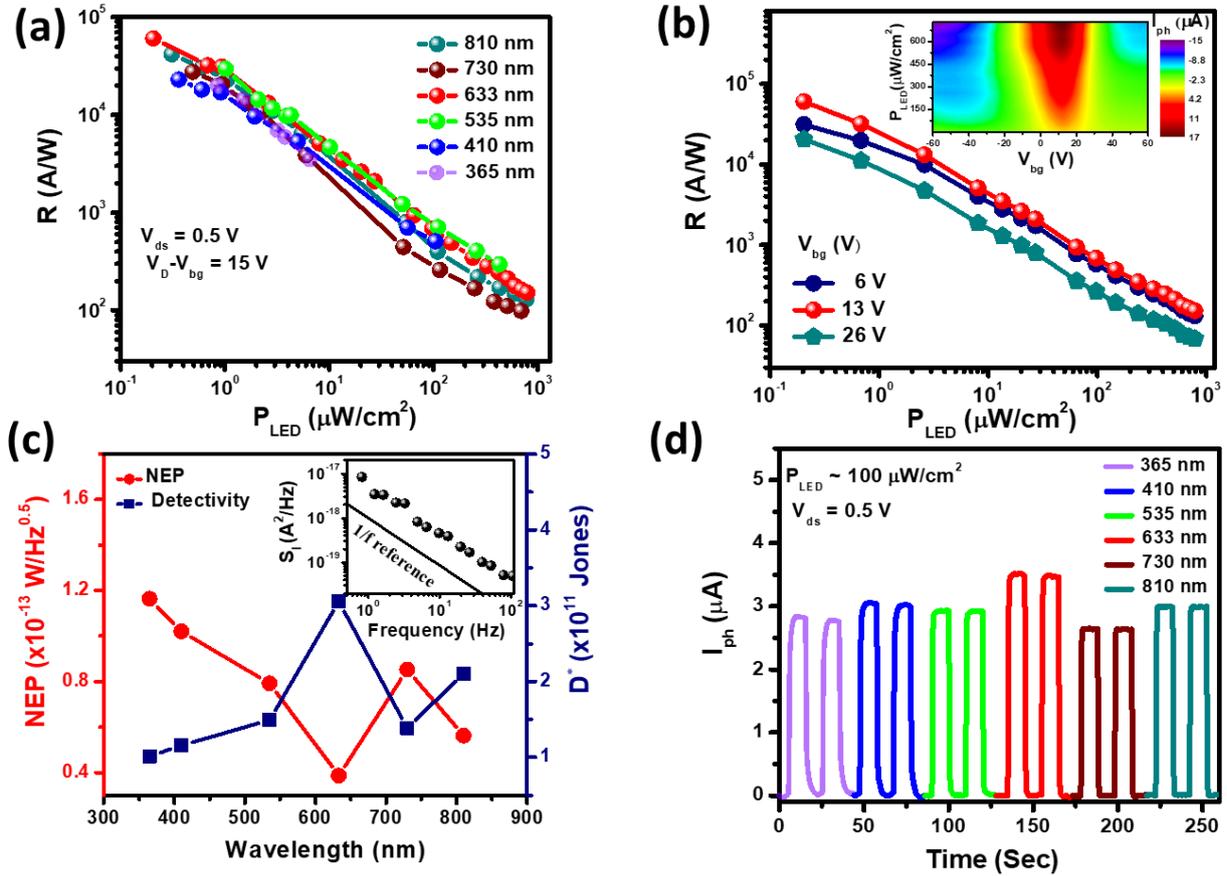

**Figure 4.** Photoresponse characteristics of graphene-MoSSe hybrid phototransistor a) Responsivity (R) as a function of illumination power ($P_{LED}$) of different wavelengths for graphene-MoSSe hybrid device at $V_{ds} = 0.5$ V, $V_D-V_{bg} = 15$ V. b) Responsivity as a function of illuminated power with different gate voltages at λ = 633 nm, $V_{ds} = 0.5$ V. 2D color plot of photocurrent ($I_{ph}$) as a function of applied gate voltages ($V_{bg}$) and illumination power ($P_{LED}$) at λ = 633 nm, $V_{ds} = 0.5$ V (Inset). c) Detectivity ($D^*$) and Noise equivalent power (NEP) as a function of wavelength with $V_{ds} = 0.5$ V, $V_D - V_{bg} = 15$ V. The inset shows the current noise density ($S_I$) of the hybrid device in dark. d) Temporal photoresponse of the hybrid device for different wavelengths at $V_{ds} = 0.5$ V, $P_{LED} \sim 100$ μW/cm², $V_{bg} = 0$ V.

The device offers a maximum responsivity of 6.06 ×10⁴ A/W for 633 nm for the minimum achievable illumination of ~ 0.2 μW/cm². In addition, a significantly higher photoresponsivity (R > 10⁴ A/W) is obtained in the wide range of UV-NIR (365-810 nm) wavelengths. The spectral photoresponse characteristics of this hybrid device is represented in **Figure S6**. This result is comparable or superior to most of the previously reported graphene-based photodetectors[46,47,48,49]. The photocurrent ($I_{ph}$) and the responsivity have a strong dependence of gate voltages ($V_{bg}$) as represented in **Figure 4b**. R increases from 3.12×10⁴ A/W to 6.06 ×10⁴ A/W when $V_{bg}$ changes from 6 V to 13 V and again decreases to 2.05×10⁴ A/W at $V_{bg}$ = 23 V. Strong gate tunability of the photocurrent is more evident for the 2D contour plot (inset,

**Figure 4b**). The gate tunable photoresponsivity and the suggested energy band diagram is shown in **Figure S7a**, **b** and **c** respectively. In addition to photoresponsivity, the noise equivalent power (NEP) and the specific detectivity (D*) are two most important figure of merits for comparing the performance of the photodetectors. NEP, i.e., the minimum optical power requires to detect the photocurrent from the noise current in a PD can be defined as[43],

$$NEP = \frac{\sqrt{S_I}}{R} \ \ \ \ \ \ \ \ \ \ \ \ \ \ \ \ \ \ \ \ \ \ \ \ \ \ \ \ \ (1)$$

Where, $S_I$ is the total noise spectral density consisting of 1/f noise, thermal noise and the shot noise[50]. The detailed calculations of $S_I$ can be found in the **Supporting information (Note 8)**. Considering the 1/f noise spectral density (**Inset, Figure 4c**), which may originate due to the trapping and detrapping of charge carriers[51], dominating the total noise power of the device, the NEP is calculated to be as low as ~$10^{-14}$ W/Hz$^{0.5}$ in our experimental wavelength range. This significantly lower NEP value indicates the ability of the device to detect weak light. Mathematically, the specific detectivity (D*) of a photodetector is defined as[43]

$$D^* = \frac{\sqrt{A}}{NEP} \ \ \ \ \ \ \ \ \ \ \ \ \ \ \ \ \ \ \ \ \ \ \ \ \ \ \ \ \ (2)$$

Where, A is the effective device area of the hybrid device. Using this equation D* is calculated as high as ~ $3.06\times10^{11}$ Jones for 633 nm wavelength. The calculated NEP and D* as a function of wavelength are shown in **Figure 4c**. **Figure 4d** presents the temporal photocurrent characteristics over the wide UV-NIR wavelength range (365-810 nm) with the same experimental condition as $V_{ds}$ = 0.5 V and $V_{bg}$ = 0 V for a constant illumination power of ~ 100 µW/cm$^2$. This hybrid graphene-MoSSe phototransistor device exhibits excellent ON/OFF modulation characteristics over a wide range of experimental wavelengths (365-810 nm) even at very low optical power and bias voltage.

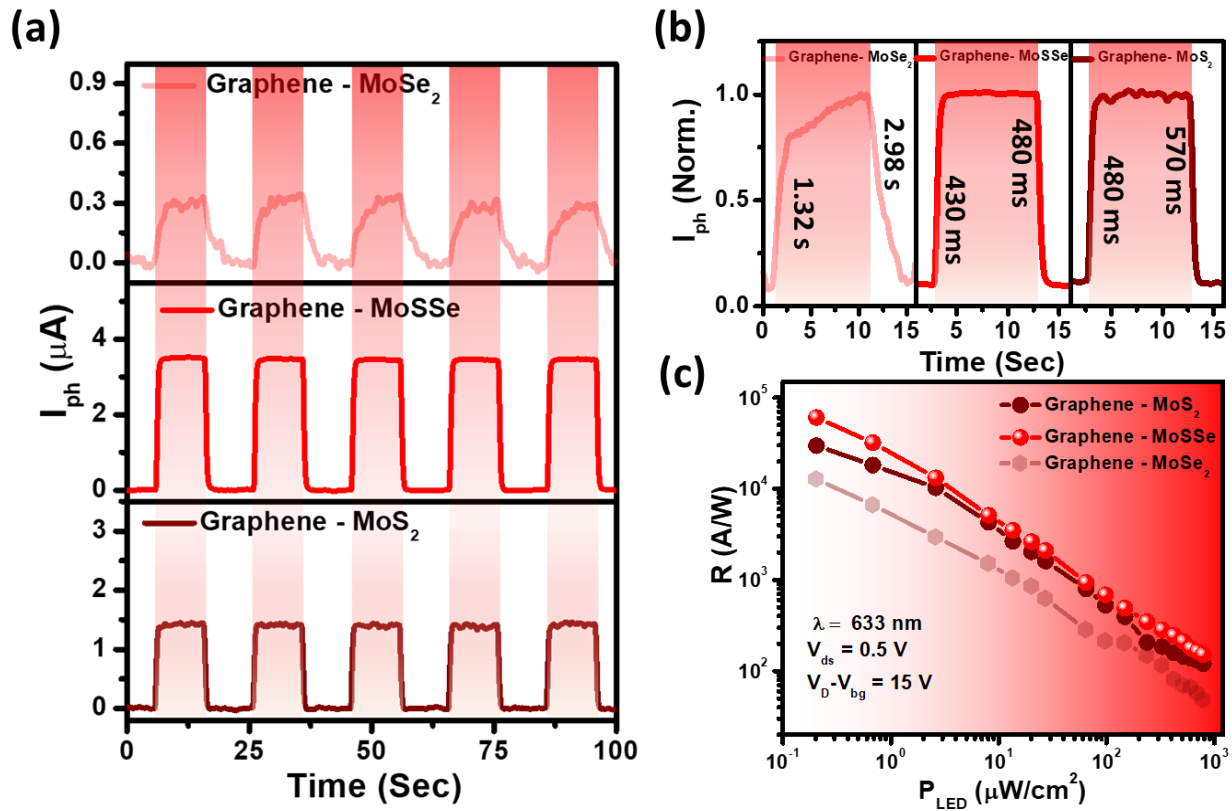

**Figure 5.** Comparison of the photoresponse between graphene MoS$_{2x}$Se$_{2(1-x)}$ hybrid devices a) Temporal photoresponse of graphene-MoS$_{2x}$Se$_{2(1-x)}$ hybrid phototransistors with x = 0, 0.5 and 1 for the same experimental conditions, $V_{ds}$= 0.5 V, $V_{bg}$= 0 V, λ = 633 nm, $P_{LED}$ ~ 100 µW/cm$^2$. b) Response time of the above hybrid phototransistors at $V_{ds}$ = 0.5 V, $V_{bg}$ = 0 V, λ = 633 nm. c) Photoresponsivity as a function of illumination power of the fabricated hybrid devices at $V_{ds}$ = 0.5 V, $V_D$-$V_{bg}$ = 15 V, λ = 633 nm.

To understand the impact of alloy engineering, it is necessary to compare the performances of three devices, i.e., graphene-MoS$_2$, graphene-MoSSe and graphene-MoSe$_2$ in identical device configurations. **Figure 5a** compares the temporal photoresponse characteristics of these three devices under the same experimental conditions of $V_{ds}$ = 0.5 V, $V_{bg}$ = 0 V, λ = 633 nm, $P_{LED}$ ~ 100 μW/cm$^2$. All the three phototransistors show robust and reproducible photocurrent in multiple ON/OFF cycles. The photocurrent ($I_{ph}$) is found to be much higher in graphene-MoSSe device compared to graphene-MoS$_2$ and graphene MoSe$_2$ based devices, indicating the superiority of MoSSe ternary alloy over its binary counterparts. The photocurrent dynamics of the devices are also evaluated by considering the time interval for the current changes from 10% to 90% (and vice versa) when light is turned on or off. From **Figure 5b**, it is seen that the rise time ($τ_{rise}$) and the fall time ($τ_{fall}$) for the graphene-MoSSe phototransistor are 430 ms and 480 ms respectively, whereas $τ_{rise}$ becomes 480 ms and 1.32 sec and $τ_{fall}$ becomes 570 ms and 2.98 sec for graphene-MoS$_2$ and graphene-MoSe$_2$ devices, respectively. The responsivity of these three phototransistors with respect to the illumination intensity are plotted in **Figure 5c** for the same experimental conditions of $V_{ds}$ = 0.5 V, $V_D$-$V_{bg}$ = 15 V, λ = 633 nm. As expected, the responsivity for the graphene-MoSSe phototransistor is much higher compared to other two devices, showing 6.06 ×10$^4$ A/W for graphene-MoSSe, 2.97×10$^4$ A/W for graphene-MoS$_2$ and 1.26×10$^4$ A/W for graphene-MoSe$_2$ at 633 nm with minimum achievable power of ~0.2 μW/cm$^2$. The NEP and D* for all the three devices under 633 nm illumination with same experimental conditions are depicted in **Figure S9.** Higher photocurrents and the ultrahigh photoresponsivity of the MoSSe alloy-based device can be explained by the conversion of deep to shallow level defect densities by minimizing the deep level defect densities, as explained previously (**Figure 3e**). The variation of photoresponsivity for all the alloy composites in the hybrid graphene-MoS$_{2x}$Se$_{2(1-x)}$ alloys is represented in **Figure S8**. **Table 1** gives the comparison of photoresponsivity (R), NEP, specific detectivity (D*) and the characteristics times of the fabricated graphene-MoS$_2$, graphene-MoSSe and graphene-MoSe$_2$ phototransistor. Stability is one of the most important parameters for designing a photodetector device since most of the materials like BP[52,53], perovskites[22,54] are very much unstable in nature. However, even without any capping layers, our ternary alloy device offers an extraordinary stability in photocurrent switching characteristics (**Figure S10**) without any degradation after 4 months of its fabrication.

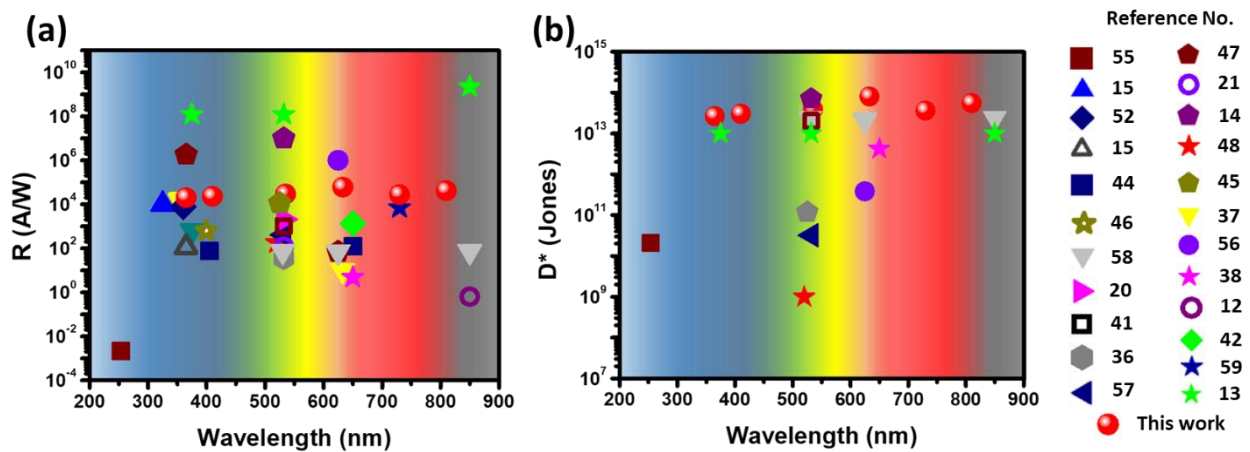

**Figure 6.** Comparison of the obtained results with some of the previously reported results a) Responsivity (R) and b) Specific detectivity (D*) comparison with similar device structure. (Considering the dark current limit only).

Finally, we compare the performances of some of the recently reported graphene-based phototransistor devices. As shown in **Figure 6a** and **6b**, the responsivity (R) and the specific detectivity ($D^*$) are summarized and compared with the previous reports. It is evident that the performance of our graphene-MoSSe phototransistor is superior to the other systems. It may be noted that the specific detectivity ($D^*$) is compared by considering the dark current only, as calculated in the reported literatures[55,56,57,58,59].

**Table 1.** Comparison of the photodetector performance parameters for all three devices i.e., graphene –$MoSe_2$, graphene – MoSSe, graphene – $MoS_2$.

| Device Name | Responsivity(R) (A/W) | NEP (W/Hz$^{0.5}$) | Detectivity ($D^*$) (Jones) | Rise Time ($\tau_{Rise}$) (Sec) | Fall Time ($\tau_{Fall}$) (Sec) |
|---|---|---|---|---|---|
| Graphene-$MoSe_2$ | $1.26 \times 10^4$ | $4.68 \times 10^{-13}$ | $2.52 \times 10^{10}$ | 1.32 | 2.98 |
| Graphene-MoSSe | $6.06 \times 10^4$ | $3.87 \times 10^{-14}$ | $3.06 \times 10^{11}$ | 0.43 | 0.48 |
| Graphene-$MoS_2$ | $2.97 \times 10^4$ | $1.30 \times 10^{-13}$ | $9.08 \times 10^{10}$ | 0.48 | 0.57 |

## Conclusions

To summarize, the composition tunable synthesis of $MoS_{2x}Se_{2(1-x)}$ ternary alloy (x = 0 to 1) nanosheets has been demonstrated using a cost effective, eco-friendly hydrothermal approach. By fabricating three terminal heterostructure device with graphene, the phototransistor action has been studied systematically with three different composites of $MoS_{2x}Se_{2(1-x)}$ (x = 0, 0.5, 1.0). The graphene-MoSSe hybrid phototransistor exhibits superior photo-sensing properties compared to the other two devices (i.e., $MoS_2$ and $MoSe_2$), which is attributed to the suppression of deep level defects and better structural stability in MoSSe. The broadband (UV-NIR) light absorption of MoSSe, ultrafast charge carrier transport in graphene and the improved interfacial property at graphene-MoSSe heterostructure make this device suitable for advanced optoelectronic devices. This graphene-MoSSe hybrid device exhibits extremely high photoresponsivity ($>10^4$ A/W), low noise equivalent power ($\sim 10^{-14}$ W/Hz$^{0.5}$), higher specific detectivity ($\sim 10^{11}$ Jones) in the wide UV-NIR (365-810 nm) wavelength range with gate tunability. Having large area scalability with bulk production of $MoS_{2x}Se_{2(1-x)}$ alloys and CVD graphene are clear advantages, these results have an important implication towards facile and scalable fabrication of high-performance optoelectronic devices based on ternary 2D alloy materials, providing an insight into the fundamental interaction between the van-der-Waals materials.

## Associated content

**Supporting information**

EDAX analysis and elemental mapping of the $MoS_{2x}Se_{2(1-x)}$ alloys, TEM and AFM images of $MoS_{2x}Se_{2(1-x)}$ alloys, characterisations of graphene-$MoS_{2x}Se_{2(1-x)}$ alloy, electrical characterisations and the calculation of noise spectral density of graphene-MoSSe hybrid phototransistor, spectral photoresponse characteristics of graphene-MoSSe phototransistor, Noise equivalent power (NEP) and the specific detectivity ($D^*$) of the graphene-based hybrid phototransistors, Stability of the hybrid graphene-MoSSe phototransistor.

## Experimental Section

**Chemicals:** Sodium molybdate dehydrate ($Na_2MoO_4 \cdot 2H_2O$), Sulfur (S) and Selenium (Se) were purchased from Alpha Aser, and Sodium borohydride ($NaBH_4$) was purchased from Sigma-aldrich. All these chemicals were directly used without further purification. Deionized water was purified using a Milli-Q system.

**Synthesis of $MoS_{2x}Se_{2(1-x)}$ Nanosheets:** Ternary alloy $MoS_{2x}Se_{2(1-x)}$ nanosheets are synthesized in hydrothermal method followed by ultrasonication. In this typical procedure $Na_2MoO_4 \cdot 2H_2O$, S and Se are used as precursors of Mo, S and Se respectively and $NaBH_4$ is used as reducing agent. For this synthesize procedure, first $Na_2MoO_4 \cdot 2H_2O$ is dissolved in 15 DI water, S and Se powder of appropriate stoichiometric amounts are added to this solution and then the resulting solution is homogenized by continuous magnetic stirring. Then aqueous $NaBH_4$ is added dropwise into the above mixture and the colour of solution is turned brown and after that the resulting solution is kept for magnetic stirring to get complete homogeneous uniform mixture. This brown coloured solution is transferred into a 40 ml teflon-lined stainless autoclave and placed in an oven at 200°C for 24 hrs. After the reaction is over, the autoclave is naturally cooled to room temperature and the finally obtained black powder samples are collected by repeated washing and centrifugation by DI water and IPA. The cleaned wet powder sample is dried and the resulting dried powder is dispersed in IPA and bath sonicated for followed by centrifugation to get the nanosheets. Finally, the ternary $MoS_{2x}Se_{2(1-x)}$ nanosheet dispersed in IPA is used for characterizations and optoelectronic experiments. For comparison, $MoS_2$ and $MoSe_2$ are prepared by similar ultrasonication assisted hydrothermal route.

**Material Characterizations:** The morphological and structural analysis of all the alloy components is carried out by using a field-emission scanning electron microscope (FESEM) equipped with an energy-dispersive X-ray (EDX) spectrometer, with an electron energy of 20 keV. To further study the detailed morphology and surface profile a high-resolution transmission electron microscope (HRTEM) (FEI-TECNAI G2 20ST, energy 200 keV) and atomic force microscopy (AFM) (di INNOVA) in tapping mode are performed. The crystallinity of the as synthesized nanosheets is investigated by X-ray diffraction (Rigaku (Smartlab)). Absorption spectrums are taken by using a UV-Vis spectrometer (Shimadzu -UV-Vis 2600 Spectrophotometer) and micro-raman and micro-photoluminescence spectrums are recorded with a spectrometer (LabRam HR Evolution; HORIBA France SAS-532 nm laser).

**Device Fabrication:** The phototransistor devices are fabricated by using commercially available single-layer CVD-grown graphene on $p^+$ doped Si/SiO$_2$ (300 nm) substrates (purchased from Graphenea, USA). Then the Ti/ Au (5 nm / 60 nm) source and drain electrodes are deposited on top of the graphene film through a shadow mask by electron beam evaporation and formed a channel of W/L = 200 µm/70 µm. After the fabrication of the back gated graphene transistor, the chemically synthesized MoS$_{2x}$Se$_{2(1-x)}$ alloys (~25 µl) are drop casted on top of the graphene transistor to make the complete hybrid devices. All the devices are then annealed at 80º C for 45 mins to remove the interfacial residues.

**Device Characterisations:** All the electrical and optical measurements are carried out at room temperature (in vacuum ~$10^{-5}$ mbar) in a homemade electronic setup having an optical window. MFLI lock in amplifier (Zurich Instruments) and a Keithley 2450 sourcemeter are used in all over the experiments with AC two-probe configuration at a carrier frequency of 226.67 Hz. For the photocurrent measurements we use well calibrated and collimated 365 nm, 410 nm, 535 nm, 633 nm, 735 nm, and 810 nm LEDs (Thorlab) sourced by DC2200 power supply with a spot size ~ 3 mm. The LED powers are calibrated by using Flame-Ocean Optics spectrometer with integrating sphere set up.

# Acknowledgements

S.M. acknowledges the INSPIRE Fellowship program, DST, Govt. of India, for providing him with a research fellowship (IF170929). The authors acknowledge the characterization facilities from the TRC project, clean room fabrication facilities of SNBNCBS and discussion with Rajib Kumar Mitra.

# Conflict of Interest

The authors declare no conflict of interest.

# Supporting Information

# High performance Broadband Photodetection Based on Graphene – MoS$_{2x}$Se$_{2(1-x)}$ Alloy Engineered Phototransistors

Shubhrasish Mukherjee[1], Didhiti Bhattacharya[1], Samit Kumar Ray[*1,2] and Atindra Nath Pal[*1]

[1]*S. N. Bose National Center for Basic Science, Sector III, Block JD, Salt Lake, Kolkata – 700106*

[2] *Indian Institute of Technology Kharagpur, 721302, West Bengal, India*

Email: physkr@phy.iitkgp.ac.in, atin@bose.res.in

**Contents**

**Supplementary Note 1:** EDAX analysis of the MoS$_{2x}$Se$_{2(1-x)}$ alloys with different composition.

**Supplementary Note 2:** EDAX elemental mapping of MoS$_{2x}$Se$_{2(1-x)}$ alloys.

**Supplementary Note 3:** TEM and AFM images of MoS$_{2x}$Se$_{2(1-x)}$ alloys with different composition.

**Supplementary Note 4:** Characterisation of the graphene – MoSSe hybrid phototransistor.

**Supplementary Note 5**: Electrical characterisation of the graphene – MoSSe hybrid phototransistor and the calculation of carrier mobility.

**Supplementary Note 6**: Spectral photoresponse characteristic of the hybrid graphene-MoSSe phototransistor.

**Supplementary Note 7**: Gate tunable photoresponse and the suggested energy band diagram of the hybrid graphene – MoSSe phototransistor.

**Supplementary Note 8:** Spectral density of noise of the hybrid graphene-MoSSe hybrid phototransistor.

**Supplementary Note 9:** Composition dependent photoresponse characteristics of the hybrid graphene-MoS$_{2x}$Se$_{2(1-x)}$ alloy phototransistors.

**Supplementary Note 10:** Noise equivalent power (NEP) and the specific detectivity (D*) of the graphene-based hybrid phototransistors with same experimental conditions.

**Supplementary Note 11:** Stability of the hybrid graphene – MoSSe phototransistor.

1. **EDAX analysis of the MoS$_{2x}$Se$_{2(1-x)}$ alloys with different composition:**

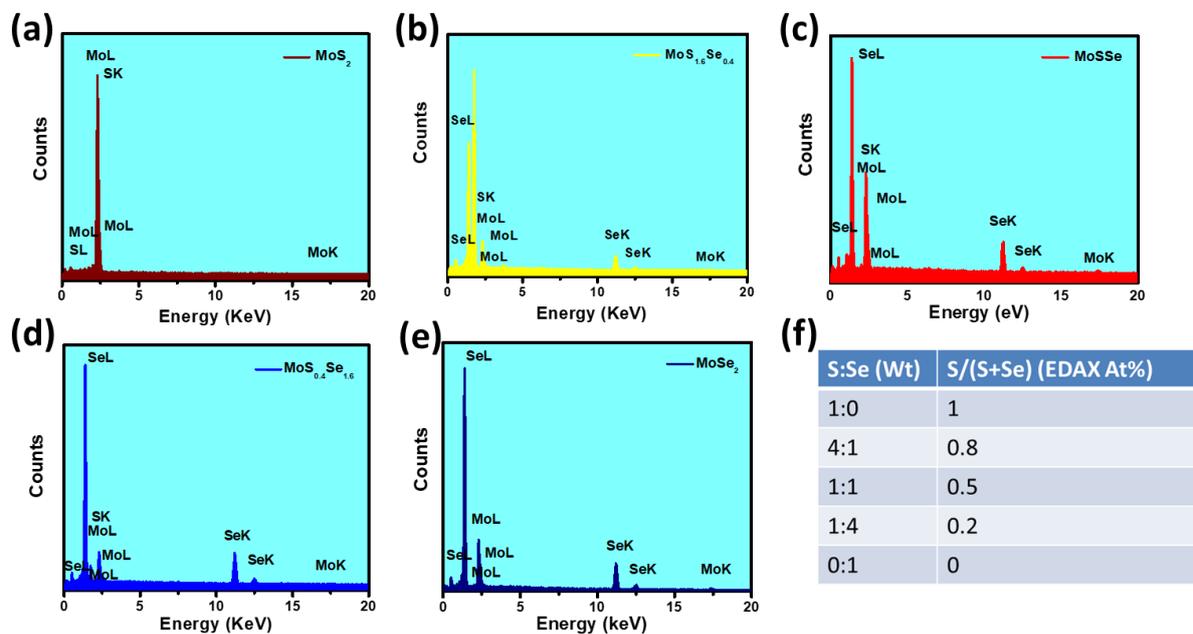

**Figure S1:** a)-e) EDAX spectra of MoS$_{2x}$Se$_{2(1-x)}$ alloys with different compositions. f) EDAX calculation table of the MoS$_{2x}$Se$_{2(1-x)}$ alloy nanosheets.

2. **EDAX elemental mapping of the MoS$_{2x}$Se$_{2(1-x)}$ alloys:**

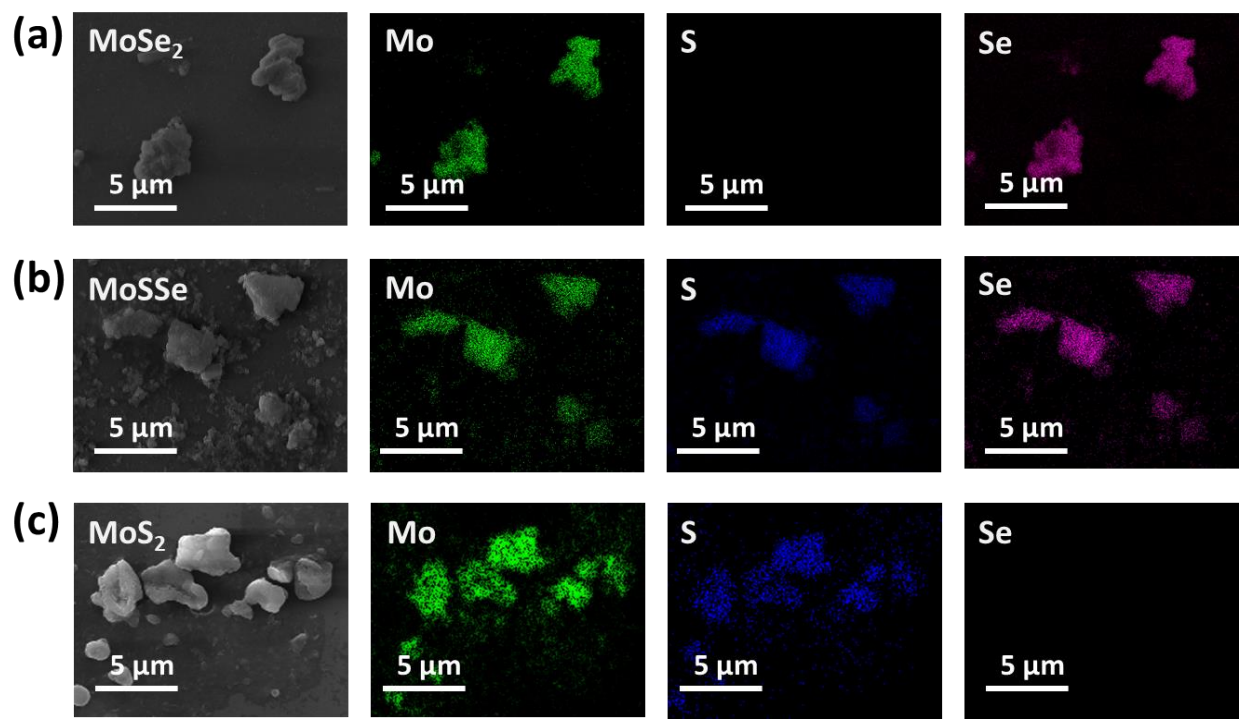

**Figure S2:** EDAX elemental mapping of different components (Mo, S, Se) of MoS$_{2x}$Se$_{2(1-x)}$ alloys a) MoSe$_2$ b) MoSSe c) MoS$_2$ nanosheets.

## 3. TEM and AFM images of MoS$_{2x}$Se$_{2(1-x)}$ alloys with different composition:

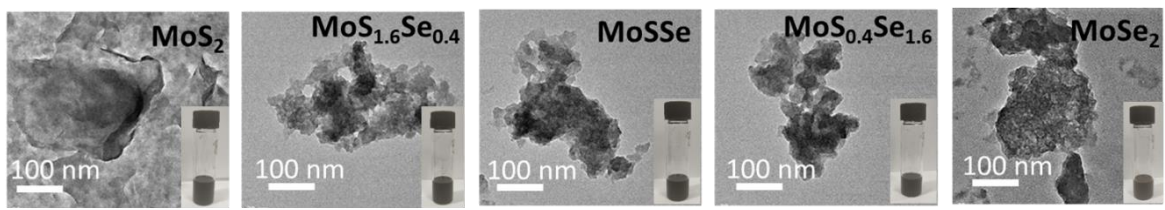

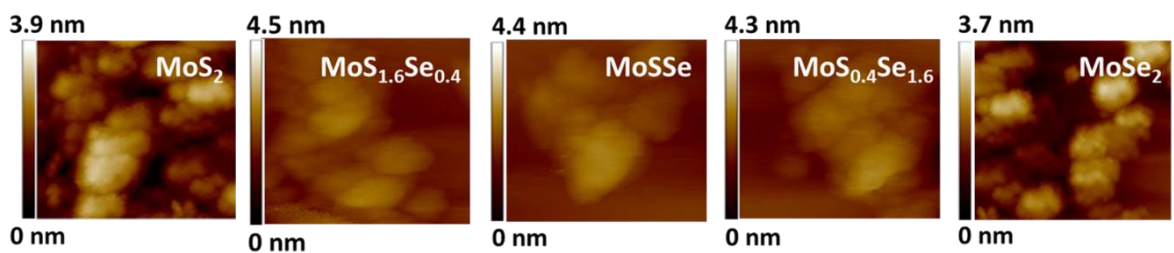

**Figure S3:** a) TEM and the optical images of dispersed nanosheets in IPA solution (Inset). b) AFM images of synthesized MoS$_{2x}$Se$_{2(1-x)}$ alloys suggesting the formation of few layered 2D sheets.

### 4. Characterisation of the graphene – MoSSe hybrid phototransistor:

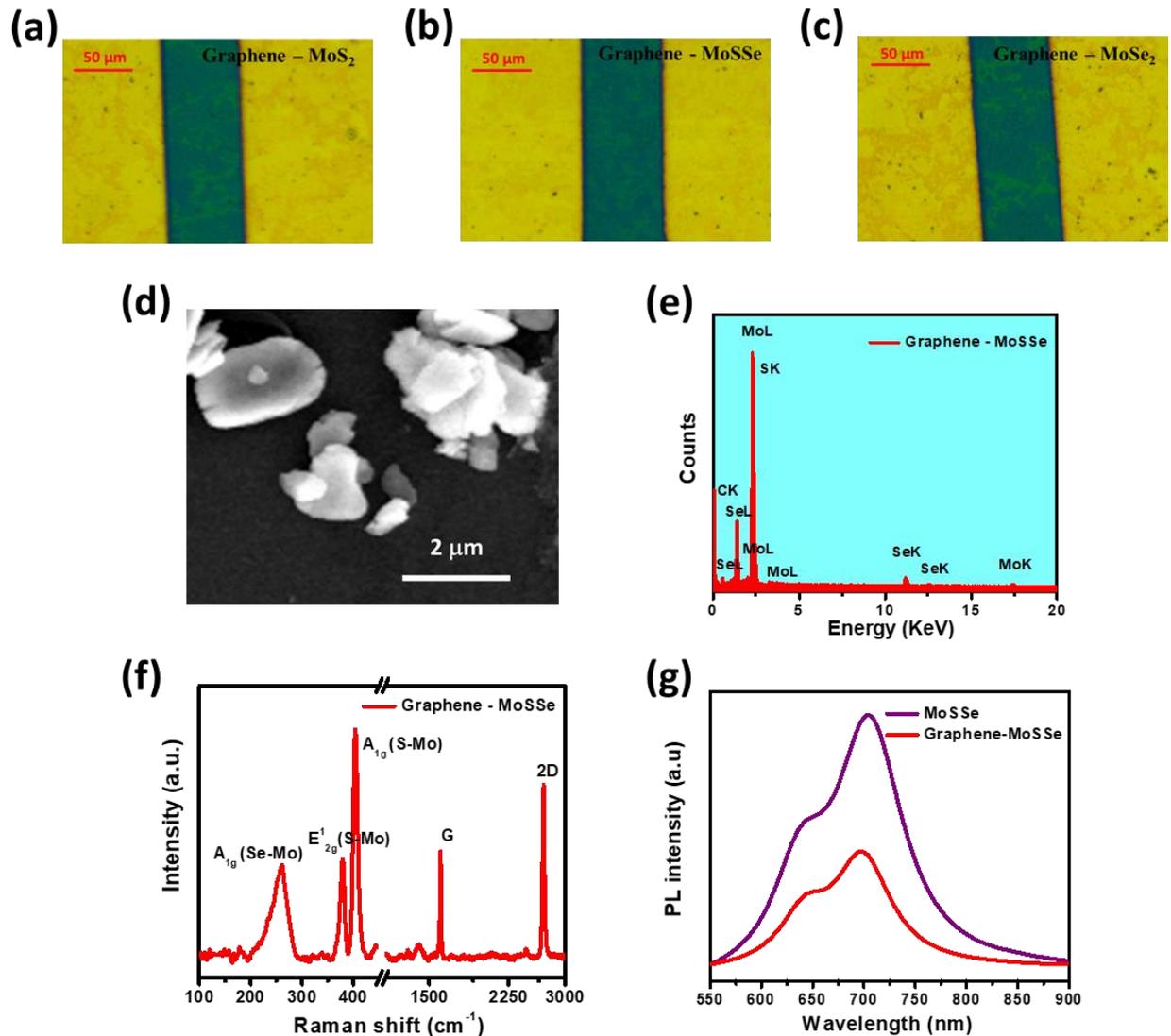

**Figure S4:** Characterizations of graphene-MoSSe hybrid phototransistor. Optical images of a) graphene- MoS$_2$, b) graphene- MoSSe, c) graphene- MoSe$_2$. d) Scanning electron micrograph (SEM) e) EDAX spectra, f) Raman spectra and g) PL spectra of bare MoSSe and the heterostructure with excitation 532 nm.

### 5. Electrical characterisation of the graphene – MoSSe hybrid phototransistor and the calculation of carrier mobility:

The field effect mobility (μ) of the hybrid graphene-WS$_2$ QDs transistor can be calculated by simply introducing Drude model which gives

$$\mu = \frac{\sigma}{ne} \quad \ldots\ldots\ldots\ldots\ldots\ldots\ldots\ldots\ldots\ldots \text{(1)}$$

Where, σ is the conductivity, n represents the charge carrier density and e is the electronic charge.

Here, σ is calculated from the relation,

$$\sigma = \frac{1}{R}\frac{L}{W} \quad \text{............................................ (2)}$$

Where, R is the resistance of the device. L and W are the effective length and width of the device Here in our device L~ 70 µm and W~ 200 µm.

The gate voltage ($V_{bg}$) is converted to carrier density (n) using a parallel plate capacitor model, i.e.

$$n = \frac{C_{bg}}{e}(V_{bg} - V_D) \quad \text{........................ (3)}$$

Where, $C_{bg} = 1.15 \times 10^{-8}$ F/cm$^2$ is the backgate capacitance of 300 nm SiO$_2$. Using the following equations, we have calculated the average field effect mobility of the charge carriers as ~ 216 cm$^2$/VS for holes and ~ 92 cm$^2$/VS for electrons in our hybrid device.

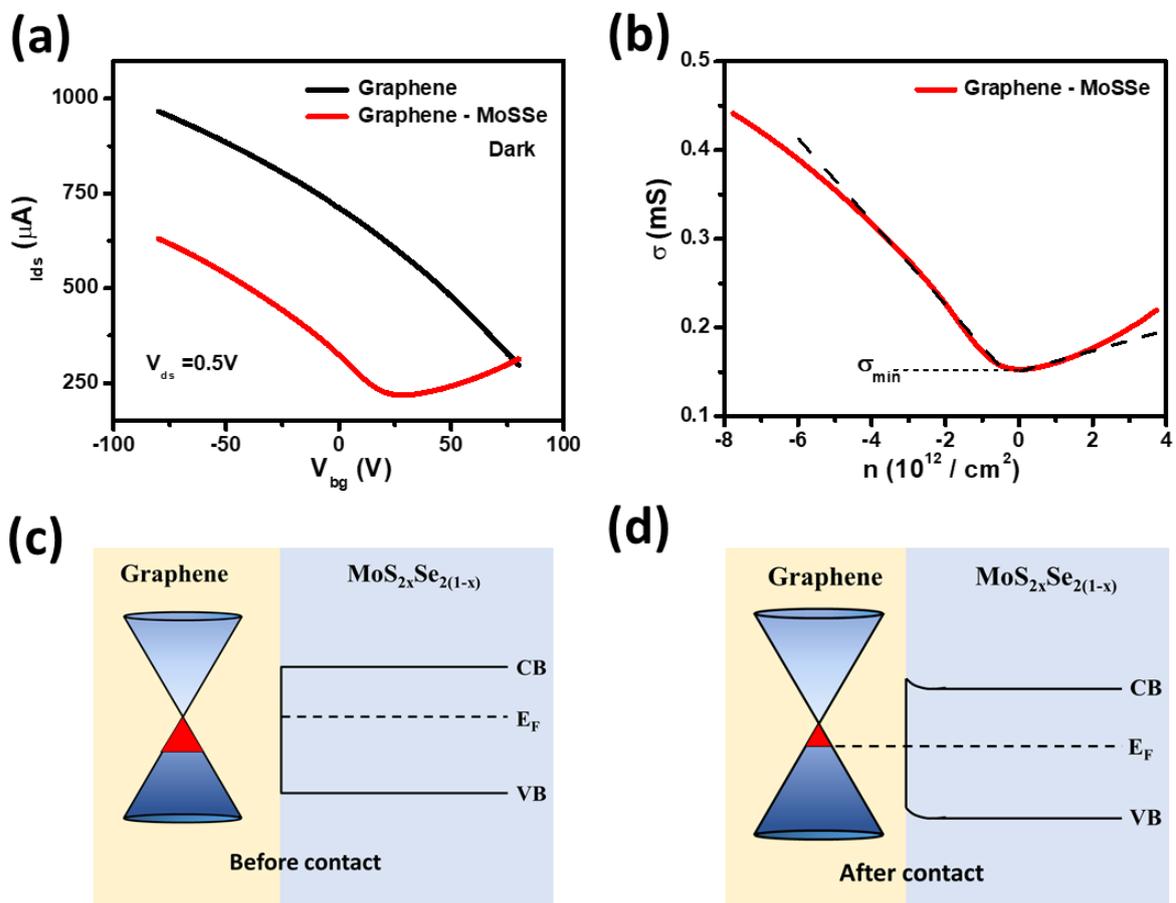

**Figure S5:** Characterizations of graphene-MoSSe hybrid phototransistor a) $I_{ds}$-$V_{bg}$ characteristics of the graphene transistor before and after deposition of MoSSe nanosheets. b) Conductivity (σ) as a function of carrier density (n) of the hybrid phototransistor. Schematic energy diagrams c) before and d) after the graphene being in contact with MoSSe alloy respectively.

## 6. Spectral photoresponse characteristic of the hybrid graphene-MoSSe phototransistor:

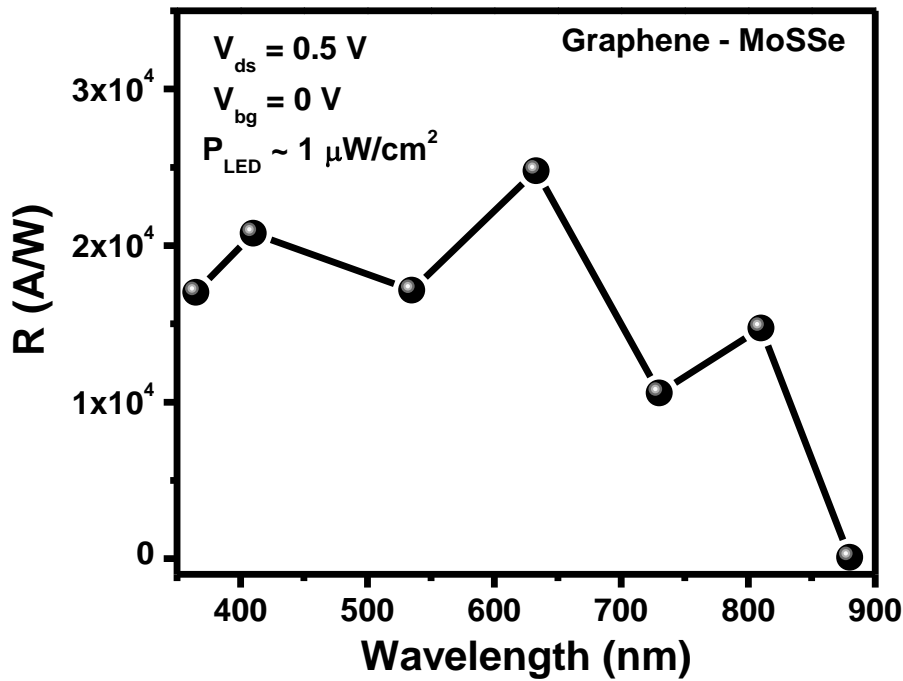

**Figure S6**: Photoresponsivity (R) as a function of wavelength of the graphene- MoSSe hybrid phototransistor under illumination of ~ 1 µW/cm² at $V_{ds}$ = 0.5 V, $V_{bg}$ = 0 V.

## 7. Gate tunable photoresponse and the suggested energy band diagram of the hybrid graphene – MoSSe phototransistor:

The dependence of photoresponssivity (R) of graphene-MoSSe hybrid phototransistor on $V_{bg}$-$V_D$ is represented in **Figure S7a**. Under illumination, electron-hole pairs are generated in MoSSe alloy and separated at the interface (due to the built-in field across graphene-MoSSe interface). For $V_{bg}$ < $V_D$, holes are the dominated charge carriers in graphene (**Figure S7b**). By reducing the gate voltages ($V_{bg}$), the Fermi level of graphene is lowered and it enhances the built-in electric field. It helps more photo-generated holes to be transferred to the graphene channel by trapping the electrons in the TMDC layer (due to the upward band bending), resulting in an increased photoresponsivity (R) (The photoresponsivity (R) attains maximum value ~ 3500 A/W at $V_{bg}$ – $V_D$ = -15 V, λ = 633 nm and $P_{LED}$ ~ 13.5 µW/cm²). With further decrease of $V_{bg}$, the interfacial barrier potential becomes thinner and the trapped photogenerated electrons (inside the TMDC layer) start tunnelling through it, resulting in a decreased photoresponsivity. Similarly, for $V_{bg} > V_D$, electrons are responsible for the conductivity in graphene. And, the increase of $V_{bg}$ increases the injection of photo-generated electrons to graphene from MoSSe due to the downward band bending in MoSSe (**Figure S7c**) resulting a negative photoresponsivity (R). The photoresponsivity becomes negligibly small near to the Dirac point ($V_D \sim 28$ V in this graphene- MoSSe device).

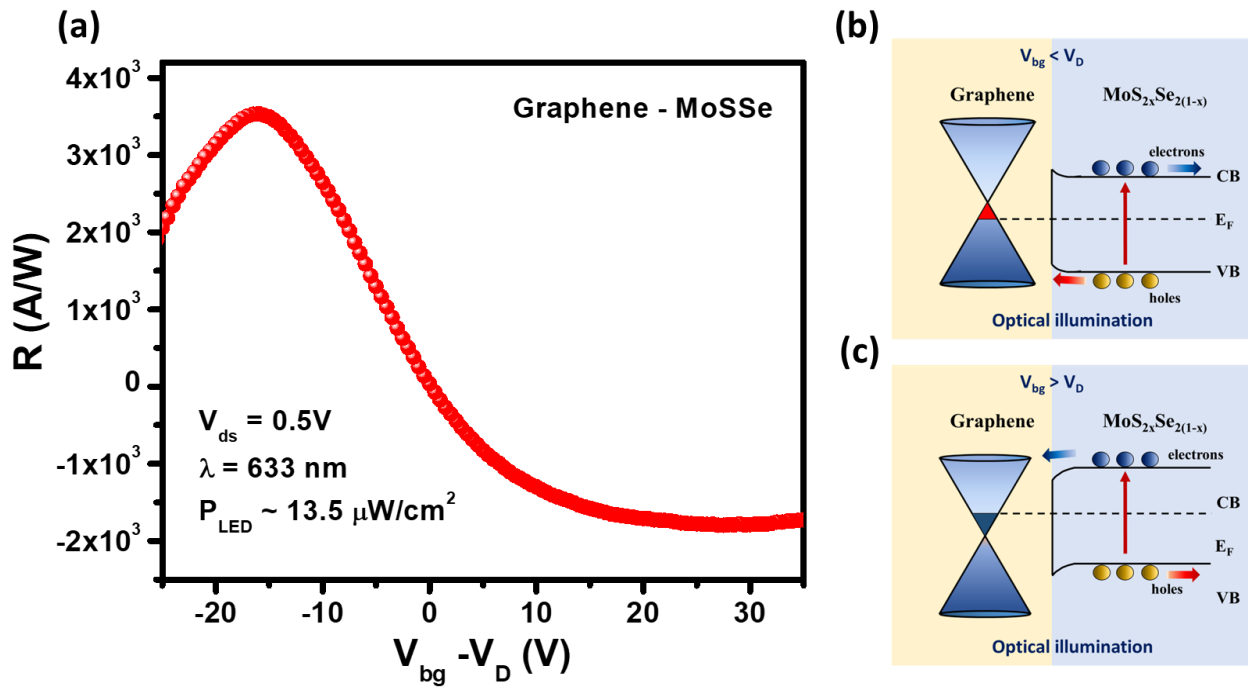

**Figure S7**: (a) Photoresponsivity (R) as a function of $V_{bg}$ of the graphene- MoSSe hybrid phototransistor under illumination of $\sim 13.5$ µW/cm$^2$ at $\lambda = 633$nm, $V_{ds} = 0.5$ V. (b) The band bending of the graphene- MoS$_{2x}$Se$_{2(1-x)}$ at $V_{bg} < V_D$ and (c) $V_{bg} > V_D$.

## 8. Spectral density of noise of the hybrid graphene-MoSSe hybrid phototransistor:

The noise equivalent power (NEP) and the specific detectivity (D$^*$) of the phototransistor can be found by considering the total spectral density of noise ($S_I$) of the device in dark current waveform. The spectral density of noise ($S_I$) for a phototransistor is defined as

$$S_I = S_I \text{(Shot)} + S_I \text{(Thermal)} + S_I (1/f) \dots\dots\dots\dots (4)$$

The spectral density of shot noise (SI (Shot)) can be calculated by the equation…

$$S_I \text{ (Shot)} = 2qI_{dark} \quad \dots\dots\dots\dots\dots\dots\dots\dots\dots\dots \quad (5)$$

Where, q is the electronic charge and $I_{dark}$ is the dark current of the hybrid device at $V_{ds} = 0.5V$ and $V_D - V_{bg} = 15$ V.

By using the above equation (equation 5), the calculated spectral density of shot noise ($S_I$ (Shot)) of the hybrid graphene – MoSSe phototransistor is $8.08 \times 10^{-23}$ A$^2$/Hz.

Also, the spectral density of thermal noise ($S_I$(Thermal)) can be calculated by the equation…

$$S_I \text{ (Thermal)} = \frac{4K_B T}{R} \quad \dots\dots\dots\dots\dots\dots\dots\dots\dots \quad (6)$$

Where, $K_B$ is the Boltzmann constant, T is the room temperature and R is the device resistance at dark. In room temperature (T = 300K), the $S_I$ (thermal) is calculated as $8.35 \times 10^{-24}$ A$^2$/Hz.

The 1/f noise spectral density ($S_I$ (1/f)) of the hybrid device is measure directly in dark with $V_{ds} = 0.5$ V, $V_D - V_{bg} = 15$ V (**Figure 4c** (Inset)). At modulation frequency 1 Hz the value of $S_I$ (1/f) is $5.5 \times 10^{-18}$ A$^2$/Hz.

From these above results it is clear that the 1/f noise dominates over the total current noise spectral density of the hybrid phototransistor device.

9. **Composition tunable photoresponse characteristics of the hybrid graphene- MoS$_{2x}$Se$_{2(1-x)}$ alloy phototransistors:**

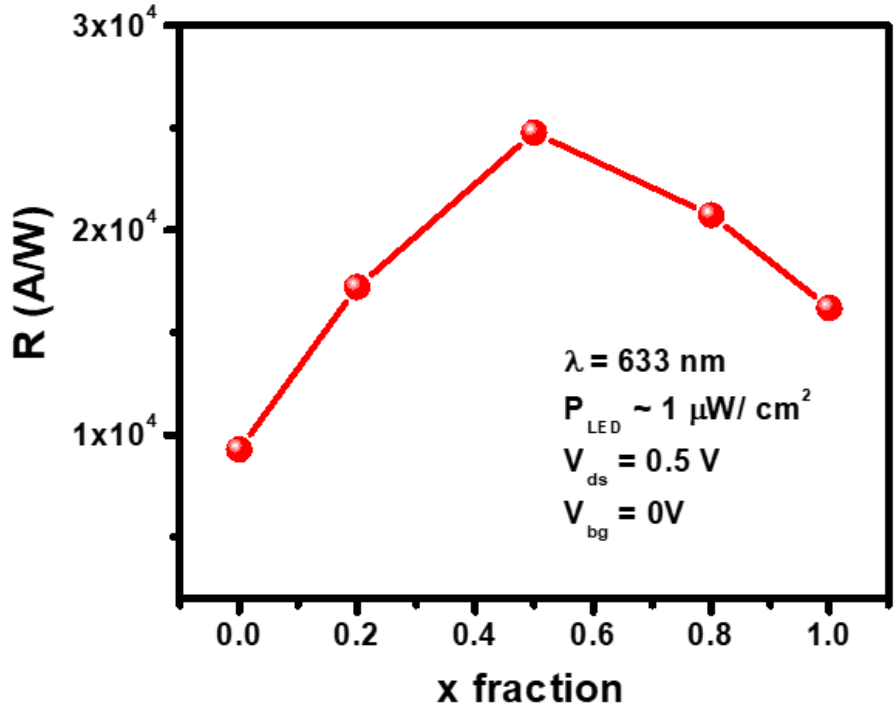

**Figure S8**: Composition (x) dependent photoresponsivity (R) hybrid graphene-$MoS_{2x}Se_{2(1-x)}$ phototransistors under illumination of ~ 1 µW/cm² at $V_{ds}$ = 0.5 V, $V_{bg}$ = 0 V.

## 10. Noise equivalent power (NEP) and the specific detectivity (D*) of the graphene-based hybrid phototransistors with same experimental conditions:

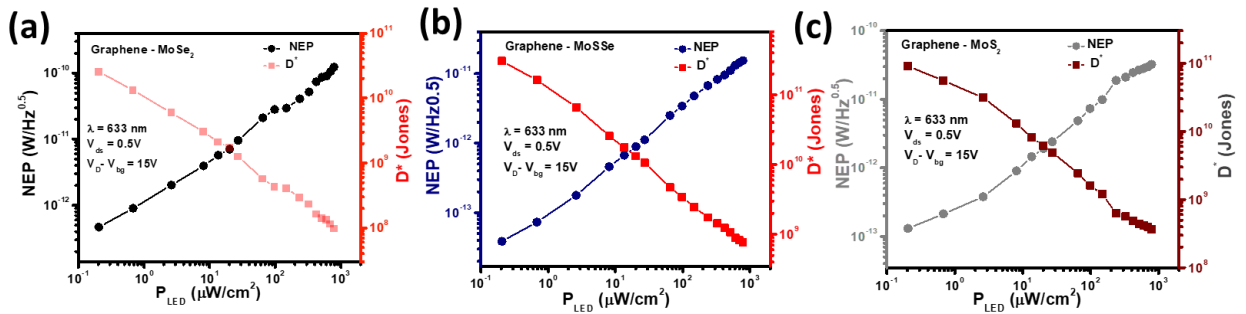

**Figure S9:** Noise equivalent power (NEP) and specific detectivity (D*) of the a) Graphene – $MoSe_2$ b) Graphene – MoSSe c) Graphene – $MoS_2$ hybrid devices as a function of illumination power ($P_{LED}$) at λ = 633 nm, $V_{ds}$ = 0.5 V and $V_D$ – $V_{bg}$ = 15 V.

## 11. Stability of the hybrid graphene – MoSSe phototransistor:

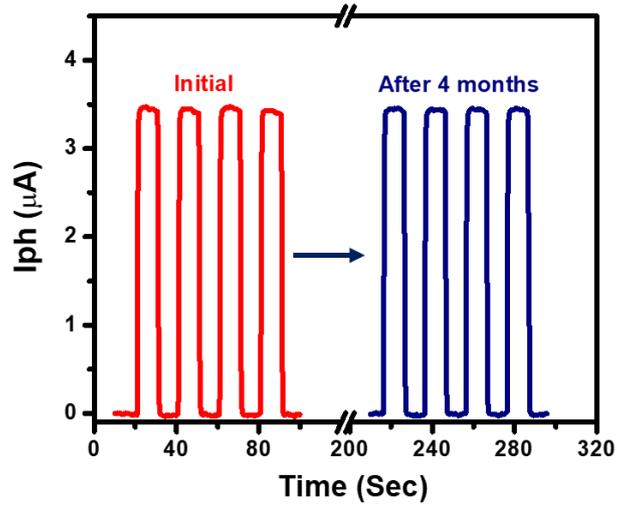

**Figure S10:** The temporal photocurrent of the graphene – MoSSe hybrid phototransistor after 4 months of its fabrication which suggests the excellent stability of the device. The experiments are performed with λ = 633 nm, $V_{ds}$ = 0.5 V, $V_{bg}$ = 0 V, $P_{LED}$ ~ 100 μW/cm$^2$.